\documentclass{rspublic}
\pdfoutput=1
\usepackage{graphicx}
\usepackage{amsmath}
\usepackage{amssymb}
\usepackage{color}

\newcommand{\revision}[1]{{#1}}

\begin{document}

\title[Multistability of free spontaneously-curved anisotropic strips]{Multistability of free spontaneously-curved anisotropic strips}

\author[L. Giomi and L. Mahadevan]{L. Giomi and L. Mahadevan}

\affiliation{School of Engineering and Applied Sciences, and Department of Physics, Harvard University, Pierce Hall 29 Oxford Street Cambridge, MA 02138, USA.\\[7pt]} 

\label{firstpage}

\maketitle

\begin{abstract}{Bistable shells; morphing structures; Gaussian curvature}
Multistable structures are  objects with more than one stable conformation, exemplified by the simple switch. Continuum versions are often elastic composite plates or shells, such as the common measuring tape or the slap bracelet, both of which exhibit two stable configurations: rolled and unrolled. Here we consider the energy landscape of a general class of multistable anisotropic strips with spontaneous Gaussian curvature. We show that while strips with non-zero Gaussian curvature can be bistable, strips with positive spontaneous curvature are always bistable, independent of the elastic moduli, strips of spontaneous negative curvature are bistable only in the presence of spontaneous twist and when certain conditions on the relative stiffness of the strip in tension and shear are satisfied. Furthermore, anisotropic strips can become tristable when their bending rigidity is small. Our study complements and extends the theory of multistability in anisotropic shells and suggests new design criteria for these structures. 
\end{abstract}

\begin{center}
	\nointerlineskip
	\rule{0.9\textwidth}{0.7pt}
\end{center}

\section{\label{sec:1}Introduction}

The notion of elastic {\em multistability} has drawn considerable attention in the past few years owing to the potential for the design of smart structures. A multistable structure is an elastic object (typically a fiber-reinforced composite) that exhibits more than one equilibrium conformation and can thus be arranged in a variety of shapes without inducing permanent deformations and with no need of mechanical hinges.  Everyday examples of these include snapping hair-clips, the slap-bracelet, and various jumping toys. Unlike more conventional engineering structures, where large deformations must be accompanied by large forces, multistable objects can switch between shapes using a small actuation force. This latter feature has made multistable structures promising candidates for the realization of a new generation of adaptive devices, in which these ``morphing'' capabilities, combined with  limited actuation, allow for switchable, controllable conformational changes. Examples of this new generation of devices are shape-changing mirrors for adaptive focusing in optical systems, morphing aircraft structures that can continuously readjust their shape to optimize aerodynamic function  (Abdulrahim {\em et al}. 2005, Mattioni {\em et al}. 2007).

The quest for a theoretical description of elastic multistability dates back to the 1920 when deployable devices, such as the measuring tape, first made their appearance. In this context, it is amusing to note that A. E. H. Love, the author of the classical treatise on elasticity (Love 1927), was once challenged to explain why the longitudinal curvature of a spring-steel measuring tape appears to match exactly its transverse curvature (Petroski 2004). He did not; indeed such an explanation was offered only decades later by Rimrott (1965) and reviewed and distilled into an elegant solution by Calladine (1988). Over the last decade the work of Iqbal {\em et al}. (2000), Galletly \& Guest (2004a, 2004b) and Guest \& Pellegrino (2006) simplified and clarified the elasticity of bistable objects, unraveling the interplay between anisotropy and spontaneous curvature in determining the shape of bistable plates. Guest \& Pellegrino (2006) used  a beam model of a strip of infinite length a finite width endowed with constant spontaneous curvature along the transverse direction. Assuming inextensibility, these authors showed that orthotropic strips exhibit a secondary equilibrium conformation in addition to the base configuration. The stability of such a secondary minimum of the elastic energy depends on the relative magnitude of the bending stiffness in the longitudinal and transverse directions as well as the twisting stiffness of the strip. Later, Seffen (2007) considered an extensible elliptical plate with free boundaries and non-zero, constant spontaneous principal curvatures and showed that even isotropic shells might be bistable.  More recently, Vidoli \& Maurini (2009) extended Seffen's uniform curvature model and showed that two-dimensional orthotropic plates with initial shallow double curvature are in fact {\em tristable} in some range of elastic moduli and spontaneous curvatures, while a one parameter family of continuously variable neutrally stable shapes was  reported by Seffen \& Guest (2011) in the context of prestressed shells.  

In this paper we complement these different results to understand the phase space for multistability in spontaneously curved elastic strips, but lift the assumptions of inextensibility and uniform curvature, with the goal of analyzing an entire range of morphing scenarios (including tristability) without compromising the simplicity of the analytical treatment. Our system consists of a free anisotropic elastic strip with spontaneous double curvature and, in general, spontaneous twist. In this setting, we look for a general solution without assuming deformations to be inextensible or the curvature to be constant across the strip. This latter feature, in particular, is what makes tristability possible even in the reduced dimensionality of strip-like plates and allow us to map a ``phase-diagram'' for the existence of bi- and tristability.   In section \ref{sec:2} we present the fundamental equations describing general anisotropic elastic strips \revision{following the approach of Mansfield (1973), Reissner (1992) and most recently, Galletly \& Guest (2004b), who considered the specific case of composite bistable tubes.}  In section \ref{sec:3} we specialize our analysis to the case of orthotropic strips and discuss the associated energy landscape. In section \ref{sec:4} we briefly discuss the case of strips with coupling between stretching and bending,  highlighting the differences and similarities with orthotropic strips. Section \ref{sec:5} concludes the paper with a focus on open problems and applications.

\section{\label{sec:2}Elasticity of anisotropic strips}

\subsection{\label{sec:2a}Formulation}

\begin{figure}[t]
\centering
\includegraphics[scale=0.7]{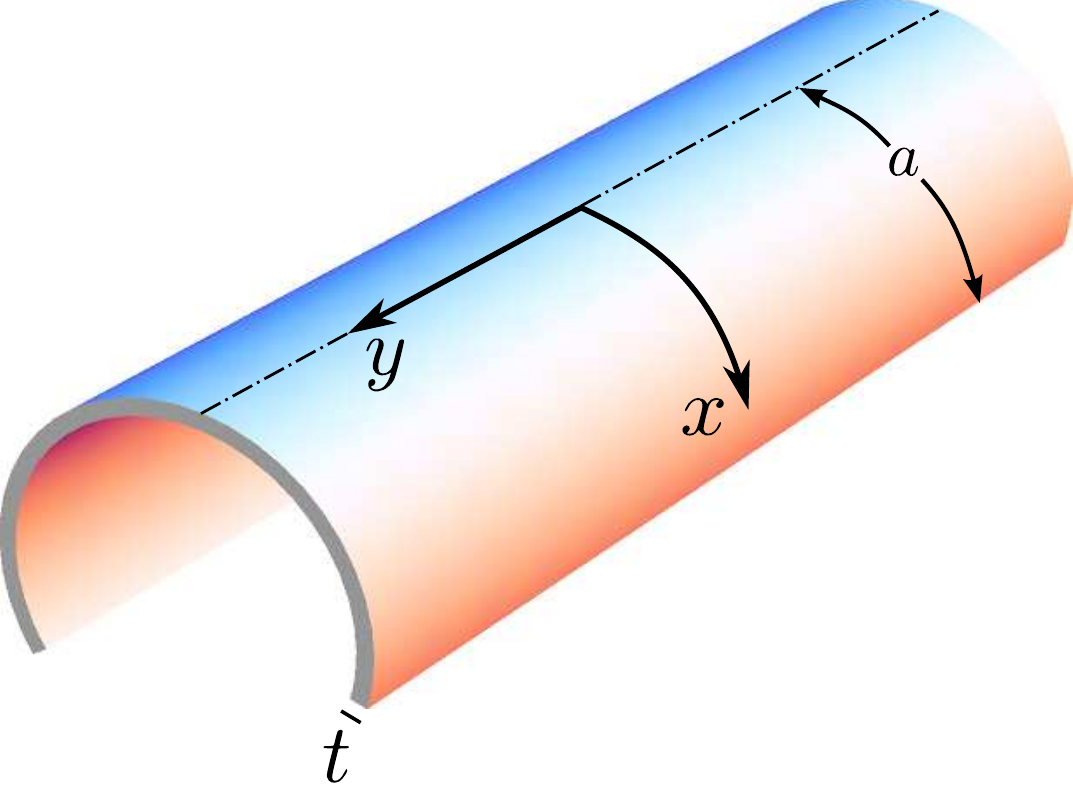}
\caption{\label{fig:strip}Schematic representation of a two-dimensional strip of thickness $t$ and width $2a$. The transverse and longitudinal directions are labelled as $x$ and $y$ respectively. For strips that are six times longer than they are wider, it is sufficient to consider them as effectively infinite along the $y$ direction.}
\end{figure}

We start by considering a long rectangular plate whose length is much larger than its width $2a$, with the $x$ the coordinate along the direction parallel to the short edge and $y$ the coordinate along the direction parallel to the long edge (Figure \ref{fig:strip}). In practice we can consider the strip infinite along the $y$ direction to avoid dealing with end conditions. As noted by Mansifeld (1973), strips with aspect ratio five or more are already well described by this approximation.  

The general constitutive equations for anisotropic plates can be expressed in terms of the ${\bf ABD}$ matrix (Ashton \& Whitney 1970; Jones 1999), that relates the in-plane stresses ${\bf N}=(N_{x},N_{y},N_{xy})$ and out-of-plane moments ${\bf M}=(M_{x},M_{y},M_{xy})$ with the strain ${\boldsymbol \epsilon}=(\epsilon_{x},\epsilon_{y},\epsilon_{xy})$ and the difference ${\boldsymbol \chi}={\boldsymbol \kappa}-{\bf c}$ between the middle-surface curvatures\footnote{We use Reissner's notation for the twisting curvature, $\kappa_{xy}=-2\partial^{2}w/\partial x \partial y$, where $w$ is the height of the middle-surface above the $xy$ plane.} ${\boldsymbol \kappa}=(\kappa_{x},\kappa_{y},\kappa_{xy})$ and their preferred (spontaneous) values ${\bf c}=(c_{x},c_{y},c_{xy})$, so that:
\begin{equation}\label{eq:constitutive-equations}
\left(
\begin{array}{c}
{\bf N} \\
{\bf M}	
\end{array}
\right) =
\left(
\begin{tabular}{c|c}
\phantom{-}{\bf A}\phantom{-} & \phantom{-}{\bf B}\phantom{-}\\
\hline
{\bf B} & {\bf D}	
\end{tabular}
\right)
\left(
\begin{array}{c}
{\boldsymbol \epsilon} \\
{\boldsymbol \chi}	
\end{array}
\right)\,,
\end{equation}
where ${\bf A}$ represents the extensional stiffness matrix, ${\bf D}$ is the bending stiffness matrix and ${\bf B}$ embodies the possible coupling between stretching and bending due to material anisotropy. 

In terms of the strain, curvature and the material properties of the shell, the elastic energy density of a general anisotropic plate is given by (Mansfield 1989):
\begin{equation}\label{eq:elastic-energy}
u = \frac{1}{2}(A_{ij}\epsilon_{i}\epsilon_{j}+2B_{ij}\chi_{i}\epsilon_{j}+D_{ij}\chi_{i}\chi_{j})
\end{equation}
where the Einstein convention on repeated indices has been used, and the first and third terms are the stretching and bending energy densities while the second term characterizes the energy associated with the coupling between bending and stretching. Since it is often useful to know the constitutive equations \eqref{eq:constitutive-equations} in their semi-inverted form, we write them as:
\begin{equation}\label{eq:semi-inverted}
\left(
\begin{array}{c}
{\boldsymbol \epsilon} \\
{\bf M}	
\end{array}
\right) =
\left(
\begin{tabular}{c|c}
$\phantom{-}{\boldsymbol \alpha}$ & $\hspace{.8ex}{\boldsymbol \beta}\phantom{-}$\\
\hline\\[-10pt]
$-{\boldsymbol \beta}^{\rm T}$ & $\,{\boldsymbol \delta}\phantom{-}$	
\end{tabular}
\right)
\left(
\begin{array}{c}
{\bf N} \\
{\boldsymbol \chi}	
\end{array}
\right)\,,
\end{equation}
where:
\begin{equation}\label{eq:elastic-moduli}
{\boldsymbol \alpha} = {\bf A}^{-1}\,,\qquad\quad
{\boldsymbol \beta } = -{\bf A}^{-1}{\bf B}\,,\qquad\quad
{\boldsymbol \delta} = {\bf D}-{\bf B}{\bf A}^{-1}{\bf B}\,.
\end{equation}
so the energy density \eqref{eq:elastic-energy} can be expressed in terms of ${\bf N}$ and $\boldsymbol{\chi}$ as:
\begin{equation}\label{eq:semi-inverted-energy-density}
u = \frac{1}{2}(\alpha_{ij}N_{i}N_{j}+\delta_{ij}\chi_{i}\chi_{j})\,.
\end{equation}
In absence of body forces or external loads, the conditions for kinematical compatibility \revision{and mechanical equilibrium} translate respectively into the following partial differential equations (Calladine 1993):
\begin{subequations}\label{eq:fvk1}
\begin{gather}
\frac{\partial^{2}\epsilon_{x}}{\partial y^{2}}-\frac{\partial^{2}\epsilon_{xy}}{\partial x\,\partial y}+\frac{\partial^{2}\epsilon_{y}}{\partial x^{2}}=-\delta K\,,\\[7pt]
\frac{\partial^{2} M_{x}}{\partial x^{2}}+2\frac{\partial^{2} M_{xy}}{\partial x\,\partial y}+\frac{\partial^{2} M_{y}}{\partial y^{2}} = \kappa_{x}N_{x}+\kappa_{xy}N_{xy}+\kappa_{y}N_{y}\,,
\end{gather}
\end{subequations}
where $\delta K=K-K_{0}$ is the difference between the Gaussian curvature of the deformed plate $K=\kappa_{x}\kappa_{y}-\kappa_{xy}^{2}$ and its spontaneous Gaussian curvature $K_{0}=c_{x}c_{y}-c_{xy}^{2}$. The radius of curvature of the strip is assumed to be much smaller than the typical length scale of the deformations (i.e. shallow shell approximation). This allows to simplify the differential structure of the theory by transforming covariant derivatives into standard partial derivatives. Eqs. \ref{eq:fvk1} are then the classic F\"oppl-von K\'arm\'an plates equations expressed in terms of stress and curvature. Since the strip is assumed to be infinitely long in the $y$ direction, all physical quantities should be invariant with respect to translations along $y$. Mechanical equilibrium requires ${\nabla \cdot}\,{\bf N}=0$, which together with the condition of translational invariance along $y$ yields:
\[
\frac{\partial N_{x}}{\partial x} = \frac{\partial N_{xy}}{\partial x} = 0\,.
\]
In absence of applied forces along the edges this implies $N_{x}=N_{xy}=0$. In addition, the Mainardi-Codazzi compatibility equations $\nabla_{i}\kappa_{jk}=\nabla_{k}\kappa_{ij}$ yield:
\[
\frac{\partial \kappa_{x}}{\partial y} -\frac{\partial \kappa_{xy}}{\partial x} = 0\,, \qquad\qquad
\frac{\partial \kappa_{y}}{\partial x} -\frac{\partial \kappa_{xy}}{\partial y} = 0\,, 
\]
that, together with the condition of translational invariance along $y$, imply that $\kappa_{y}$ and $\kappa_{xy}$ are both constant. Thus, equations \eqref{eq:fvk1} simplify to:
\begin{subequations}\label{eq:fvk2}
\begin{gather}
\frac{\partial^{2}\epsilon_{y}}{\partial x^{2}}=-\delta K\,,\\[7pt]
\frac{\partial^{2} M_{x}}{\partial x^{2}} = \kappa_{y}N_{y}\,.
\end{gather}
\end{subequations}
Using the semi-inverted constitutive equations \eqref{eq:semi-inverted}, one can easily rewrite \eqref{eq:fvk2} in the form of a single differential equation for the transverse moment $M_{x}$. To do this we start by writing:
\begin{subequations}\label{eq:fvk3}
\begin{gather}
\epsilon_{y} = \alpha_{22}N_{y}+\beta_{21}\chi_{x}+\beta_{22}\chi_{y}+\beta_{26}\chi_{xy}\,,\\[7pt]
M_{x} = -\beta_{21}N_{y}+\delta_{11}\chi_{x}+\delta_{12}\chi_{y}+\delta_{16}\chi_{xy}\,.
\end{gather}
\end{subequations}
Then, inverting (\ref{eq:fvk3}b) and using (\ref{eq:fvk2}b) we get:
\begin{subequations}\label{eq:curvature-equation}
\begin{gather}
\kappa_{x} = c_{x}+\frac{1}{\delta_{11}}\left(M_{x}+\frac{\beta_{21}}{\kappa_{y}}\frac{\partial^{2}M_{x}}{\partial x^{2}}-\delta_{12}\chi_{y}-\delta_{16}\chi_{xy}\right)\,,\\[7pt]
\frac{\partial^{2}\kappa_{x}}{\partial x^{2}} = \frac{1}{\delta_{11}}\left(\frac{\partial^{2} M_{x}}{\partial x^{2}}+\frac{\beta_{21}}{\kappa_{y}}\frac{\partial^{4}M_{x}}{\partial x^{4}}\right)\,.
\end{gather}
\end{subequations}
Combining these two equations with \eqref{eq:fvk2} we finally obtain:
\begin{multline}\label{eq:fvk4}
(\alpha_{22}\delta_{11}+\beta_{21}^{2})\,\frac{\partial^{4}M_{x}}{\partial x^{4}}+2\beta_{21}\kappa_{y}\frac{\partial^{2}\,M_{x}}{\partial x^{2}}+\kappa_{y}^{2}M_{x}\\[7pt]
= \kappa_{y}\left\{\delta_{11}(c_{x}c_{y}-c_{xy}^{2}+\kappa_{xy}^{2})+\kappa_{y}[\delta_{12}(\kappa_{y}-c_{y})+\delta_{16}(\kappa_{xy}-c_{xy})-\delta_{11}c_{x}]\right\}\,.\\
\end{multline}
The boundary conditions associated with equations \eqref{eq:fvk2} for a strip with no applied forces and torques at the lateral edges have been discussed by Reissner (1993) and translate into the requirement:
\begin{equation}\label{eq:boundary-condition}
M_{x} = \frac{\partial M_{x}}{\partial x} = 0\qquad x=\pm a\,.
\end{equation}
Together, \eqref{eq:fvk4} and \eqref{eq:boundary-condition} complete the formulation of the boundary value problem for the behavior of a long anisotropic strip with spontaneous curvature and twist.

\subsection{\label{sec:2b}Anisotropy classes}

Before we go further, it is useful to review the different form of elastic anisotropy, following the theory of laminates (Jones 1999).  Various forms of matrix coupling the in-plane strain to the out-of-plane curvature ${\bf B}$ can be obtained by controlling the relative orientation of the laminae forming the layers of a composite plate. Thus in a cross-ply fiber reinforced composite, in which the fibers in each layer are alternatively oriented at 0$^\circ$ and 90$^\circ$ with respect to the $y$ axis of the strip, one has:
\begin{equation}
{\bf B} = 
\left(
\begin{array}{ccc}
B_{11} & 0  & 0 \\
0 & -B_{11} & 0 \\
0 & 0 & 0    	
\end{array}
\right)\,.
\end{equation}
with $B_{11}=E_{\perp}h^{2}(E_{\parallel}/E_{\perp}-1)/4P$, where $E_{\parallel}$ and $E_{\perp}$ are the Young moduli in the parallel and transverse directions of the fiber and $P$ is the number of layers of the laminate. Antisymmetric angle-ply laminates, on the other hand, have laminae oriented as same angle $\theta$ with respect to the laminate coordinate axis on one side of the middle-surface and the corresponding equal thickness laminae oriented at $-\theta$ on the other side at the same distance form the middle-surface. In this case:
\begin{equation}
{\bf B} = 
\left(
\begin{array}{ccc}
0 & 0 & B_{16} \\
0 & 0 & B_{26} \\
B_{16} & B_{26} & 0    	
\end{array}
\right)\,.
\end{equation}
General asymmetric laminates obtained by the asymmetric stacking of isotropic layers with different material properties about the middle-surface, have finally:
\begin{equation}
{\bf B} = 
\left(
\begin{array}{ccc}
B_{11} & B_{12} & 0 \\
B_{12} & B_{22} & 0 \\
0 & 0 & B_{66}    	
\end{array}
\right)\,.
\end{equation}
In the next section we start with a focus on the case of orthotropic strips for which $B_{ij}=0$, while the case of angle-ply laminates is discussed in Section \ref{sec:4} in the context of tristable strips.

\subsection{\label{sec:2c}Equilibrium conformations}

In the case of naturally flat strips with no coupling between stretching and bending (i.e. $B_{ij}=0$ and $\delta_{ij}=D_{ij}$), equation \eqref{eq:fvk4} reduces to the equation given by Reissner (1993) for the case of a inhomogeneous anisotropic strip:
\[
\alpha_{22}\,\frac{\partial^{4}M_{x}}{\partial x^{4}}+\frac{\kappa_{y}}{D_{11}}\,M_{x} = \kappa_{y}\,\frac{D_{12}\kappa_{y}^{2}+D_{16}\kappa_{y}\kappa_{xy}+D_{11}\kappa_{xy}^{2}}{D_{11}}\,.
\] 
Now, letting:
\begin{gather*}
\varphi_{1} = -\frac{\beta_{21}\kappa_{y}}{\alpha_{22}\delta_{11}+\beta_{21}^{2}}\,, \qquad
\varphi_{2} =  \frac{\kappa_{y}^{2}}{\alpha_{22}\delta_{11}+\beta_{21}^{2}}
\end{gather*}
and:
\[
\varphi_{3} =  \frac{\kappa_{y}\left\{\delta_{11}(c_{x}c_{y}-c_{xy}^{2}+\kappa_{xy}^{2})+\kappa_{y}[\delta_{12}(\kappa_{y}-c_{y})+\delta_{16}(\kappa_{xy}-c_{xy})-\delta_{11}c_{x}]\right\}}{\alpha_{22}\delta_{11}+\beta_{21}^{2}}\,,	
\]
the solution of equation \eqref{eq:fvk4} which satisfies the homogeneous boundary conditions \eqref{eq:boundary-condition} is given by:
\begin{equation}\label{eq:moment}
M_{x} = -\delta_{11}(\kappa-\kappa_{0}) + \delta_{11}C_{1}\cosh k_{1}x\cos k_{2}x + \delta_{11}C_{2}\sinh k_{1}x\sin k_{2}x\,,	
\end{equation}
where:
\begin{subequations}\label{eq:kappas}
\begin{gather}
\kappa_{0} = \frac{(c_{x}c_{y}-c_{xy}^{2})+\kappa_{xy}^{2}}{\kappa_{y}}\,,\\
\kappa = c_{x}+\frac{\delta_{12}}{\delta_{11}}(c_{y}-\kappa_{y})+\frac{\delta_{16}}{\delta_{11}}(c_{xy}-\kappa_{xy})\,. 	
\end{gather}
\end{subequations} 
and:
\begin{subequations}\label{eq:constants}
\begin{gather}
C_{1} = (\kappa-\kappa_{0})\,\left(\frac{k_{1}\cosh ak_{1} \sin ak_{2}+k_{2}\sinh ak_{1} \cos ak_{2}}{k_{1}\sin ak_{2} \cos ak_{2}+ k_{2} \sinh ak_{1}\cosh ak_{1}}\right)\,,\\[10pt]
C_{2} = (\kappa-\kappa_{0})\,\left(\frac{k_{2}\cosh ak_{1} \sin ak_{2}-k_{1}\sinh ak_{1} \cos ak_{2}}{k_{1}\sin ak_{2} \cos ak_{2}+ k_{2} \sinh ak_{1}\cosh ak_{1}}\right)\,, 
\end{gather}
\end{subequations}
with:
\[
k_{1} = \left(\frac{\sqrt{\varphi_{2}}+\varphi_{1}}{2}\right)^{\frac{1}{2}}\,, \qquad
k_{2} = \left(\frac{\sqrt{\varphi_{2}}-\varphi_{1}}{2}\right)^{\frac{1}{2}}\,,
\]
The longitudinal stress $N_{y}$ and the transverse curvature $\kappa_{x}$ can be readily calculated by replacing $M_{x}$ in equations (\ref{eq:fvk2}b) and (\ref{eq:curvature-equation}a). This yields:
\begin{equation}\label{eq:anisotropic-stress}
N_{y} = \delta_{11}C_{2}'\cosh k_{1}x\cos k_{2}x -\delta_{11}C_{1}'\sinh k_{1}x\sin k_{2}x\,,
\end{equation}	
with:
\[
C_{1}'= \frac{(\alpha_{22}\delta_{11})^{\frac{1}{2}}C_{1}+\beta_{21}C_{2}}{\alpha_{22}\delta_{11}+\beta_{21}^{2}}\,, \qquad
C_{2}'= \frac{(\alpha_{22}\delta_{11})^{\frac{1}{2}}C_{2}-\beta_{21}C_{1}}{\alpha_{22}\delta_{11}+\beta_{21}^{2}}\,,
\]
and finally:
\begin{equation}\label{eq:anisotropic-curvature}
\kappa_{x} = \kappa_{0}+(C_{1}+\beta_{21}C_{2}')\cosh k_{1}x \cos k_{2}x + (C_{2}-\beta_{21}C_{1}') \sinh k_{1}x \sin k_{2}x\,.	
\end{equation}
Eq. \eqref{eq:anisotropic-curvature} is equivalent to an expression given by Galletly \& Guest (2004b) for the case of a bistable composite slit tube. We will now examine the effect of spontaneous curvature and elastic anisotropy on the energy landscape of elastic strips. 

\section{\label{sec:3}Multistable configurations of orthotropic strips}

As a starting point in our analysis we consider the case of orthotropic strips. For this class of materials there is no coupling between stretching and bending, hence $B_{ij}=0$ and the ${\bf A}$ and ${\bf D}$ matrices can be expressed in the form:
\[
{\bf A} = 
tE_{0}
\left[
\begin{array}{ccc}	
1 & \nu & 0 \\
\nu & \beta & 0 \\
0 & 0 & \rho\left(1-\frac{\nu^{2}}{\beta}\right)
\end{array}
\right]\,,
\qquad
{\bf D} = 
\frac{t^{3} E_{0}}{12}
\left[
\begin{array}{ccc}	
1 & \nu & 0 \\
\nu & \beta & 0 \\
0 & 0 & \rho\left(1-\frac{\nu^{2}}{\beta}\right)
\end{array}
\right]\,,
\]
where $E_{0}=E/(1-\nu^{2}/\beta)$ and, following Seffen (2007), we have called:
\[
E_{x} = E\,,\qquad
E_{y} = \beta E\,,\qquad
G = \rho E\,,\qquad
\nu_{yx} = \nu\,, \qquad
\nu_{xy} = \frac{\nu}{\beta}\,,
\]
where $E_{x}$ and $E_{y}$ are the Young's moduli in the $x$ and $y$ direction, $G$ is the shear modulus and $\nu_{xy}$ and $\nu_{yx}$ the Poisson ratios. The conditions for the matrices ${\bf A}$ and ${\bf D}$ to be positive definite, translate into the requirement $\beta>\nu^{2}$ and $\rho>0$ (Vidoli \& Maurini, 2009). The transverse curvature \eqref{eq:anisotropic-curvature} and the longitudinal stress \eqref{eq:anisotropic-stress} then simplify to yield:
\begin{subequations}\label{eq:orthotropic-solution}
\begin{gather}
\kappa_{x} = \kappa_{0}+C_{1}\cosh kx\cos kx+C_{2}\sinh kx\sin kx\,, \\[7pt]
N_{y} = (t\beta ED)^{\frac{1}{2}}(C_{2}\cosh kx\cos kx-C_{1}\sinh kx\sin kx)\,,
\end{gather}
\end{subequations}
where $D=t^{3}E/12(1-\nu^{2}/\beta)$, $k_{1}=k_{2}=k$ and:
\[
k =  \left(\frac{t\beta E}{4D}\,\kappa_{y}^{2}\right)^{\frac{1}{4}}\,.
\] 
The stretching energy density $u_{s}$ and the bending energy density $u_{b}$ are obtained by substituting \eqref{eq:orthotropic-solution} into \eqref{eq:semi-inverted-energy-density} and yields:
\begin{subequations}\label{eq:energy-orthotropic}
\begin{gather}
u_{s} =	\frac{1}{2t\beta E}\,N_{y}^{2}\,, \\
u_{b} = \frac{1}{2}D\,\left[\chi_{x}^{2}+\beta\chi_{y}^{2}+2\nu\chi_{x}\chi_{y}+\rho\left(1-\frac{\nu^{2}}{\beta}\right)\chi_{xy}^{2}\right]\,.
\end{gather}
\end{subequations}
Integrating \eqref{eq:energy-orthotropic} along the width of the strips we obtain the total stretching energy per unit length:
\begin{equation}\label{eq:fs}
U_{s} = \frac{1}{2} D(\kappa-\kappa_{0})^{2}\left(\frac{\Psi}{2k}-2a\Phi\right)\,,
\end{equation}
where:
\[
\Psi = \frac{\cosh 2ka - \cos 2ka}{\sinh 2ka + \sin 2ka}\,,\qquad
\Phi = \frac{\sinh 2ka \sin 2ka}{(\sinh 2ka + \sin 2ka)^{2}}\,,
\]
and the bending energy
\begin{multline}\label{eq:fb}
U_{b} 
= aD \bigg[(\kappa_{0}-c_{x})^{2}+\beta(\kappa_{y}-c_{y})^{2}+2\nu(\kappa_{0}-c_{x})(\kappa_{y}-c_{y})\\
+ \rho\left(1-\frac{\nu^{2}}{\beta}\right)(\kappa_{xy}-c_{xy})^{2}
+ \frac{\eta}{2ka}(\kappa-\kappa_{0})\Psi
+ (\kappa-\kappa_{0})^{2}\Phi
\bigg]\,,
\end{multline}
where:
\begin{equation}
\eta = \frac{5}{2\kappa_{y}}\left[(\kappa_{y}-c_{y})(\nu \kappa_{y}-c_{x})+(\kappa_{xy}^{2}-c_{xy}^{2})\right]\,,
\end{equation}
Before proceeding with the analysis, it is useful to introduce a set of dimensionless quantities defined by:
\[
\hat{x} = \frac{x}{a}\,,\qquad
\hat{t} = \frac{t}{a}\,,\qquad
\hat{c}_{i} = ac_{i}\,,\qquad
\hat{\kappa}_{i} = a\kappa_{i}\,\qquad
(i = x,\,y,\,xy\,).
\]
so that the dimensionless stresses and energies per unit length are given by:
\[
\hat{N}_{y} = \frac{N_{y}}{aE}\,, \qquad 
\hat{U}_{s} = \frac{U_{s}}{taE}\,,\qquad
\hat{U}_{b} = \frac{U_{b}}{taE}\,,\qquad
\hat{U} = \hat{U}_{s}+\hat{U}_{b}\,.
\]
$\hat{U}$ is thus the total elastic energy of a strip that accounts for both stretching and bending. In the following sections we will analyze the ``energy landscape'' embodied in $\hat{U}$  and given by \eqref{eq:fs} and \eqref{eq:fb} in a variety of scenarios. We will start by considering configurations of constant transverse curvature $\hat{\kappa}_{x}$ and then move on the more general case in which $\hat{\kappa}_{x}$ is allowed to vary across the strip. In section (\ref{sec:3c}) we will examine the limit of vanishing bending stiffness and show how tristability arises in this setting for a broad range of spontaneous curvatures.

\subsection{\label{sec:3a}Uniform curvature configurations}

To understand the energy landscape of the strip, we analyze the total elastic energy $\hat{U}$ given by \eqref{eq:fs} and \eqref{eq:fb}. We first focus on configurations having constant transverse curvature $\hat{\kappa}_{x}$: \revision{if $\hat{\kappa}_{y}$ is nonzero}, from equation \eqref{eq:constants} and (\ref{eq:orthotropic-solution}a) it follows that $C_{1}=C_{2}=0$ and $\hat{\kappa}=\hat{\kappa}_{0}$. This solution has zero longitudinal stress and, as a consequence, is an isometry of the base configuration: a deformation that preserves the local metric of the shell so that $\boldsymbol{\epsilon}=\boldsymbol{0}$. The latter statement can be also verified by calculating the Gaussian curvature of the deformed strip. Since $\hat{\kappa}_{x}=\hat{\kappa}=\hat{\kappa}_{0}$, equation (\ref{eq:kappas}a) yields:
\begin{equation}\label{eq:isometry}
\hat{\kappa}_{x} = \frac{(\hat{c}_{x}\hat{c}_{y}-\hat{c}_{xy}^{2})+\hat{\kappa}_{xy}^{2}}{\hat{\kappa}_{y}}\,,
\end{equation}
which implies that $\hat{\kappa}_{x}\hat{\kappa}_{y}-\hat{\kappa}_{xy}^{2}=\hat{c}_{x}\hat{c}_{y}-\hat{c}_{xy}^{2}$ as required by Gauss' {\em theorema egregium} (Kreyszig 1991). The corresponding total elastic energy per unit length $\hat{U}$ is given by:
\begin{equation}
\hat{U} = \frac{\hat{t}^{2}}{12}\,[\beta(\hat{\kappa}_{y}-\hat{c}_{y})^{2}+\rho(\hat{\kappa}_{xy}-\hat{c}_{xy})^{2}]\,.
\end{equation}
The condition $\hat{\kappa}=\hat{\kappa}_{0}$ used to obtain the isometric solution \eqref{eq:isometry} describes, in the plane $(\hat{\kappa}_{y},\hat{\kappa}_{xy})$ a continuous set of configurations of zero stretching energy. The explicit form of this geometrical locus is given by:
\begin{equation}
\nu\hat{\kappa}_{y}^{2}+\hat{\kappa}_{xy}^{2}-(\hat{c}_{x}+\nu \hat{c}_{y})\hat{\kappa}_{y}=\hat{c}_{xy}^{2}-\hat{c}_{x}\hat{c}_{y}\,,
\end{equation}
that corresponds to an ellipse with focii at :
\[
\hat{\kappa}_{y,\,0} = \frac{1}{2\nu}\,(\hat{c}_{x}+\nu\hat{c}_{y})\,,\qquad
\hat{\kappa}_{xy,\,0} = 0\,,
\]
and semi-axes $a_{y}=1/\ell\sqrt{\nu}$ and $a_{xy}=1/\ell$ with $\ell^{-2}=\frac{1}{4\nu}(\hat{c}_{x}-\nu \hat{c}_{y})^{2}+\hat{c}_{xy}^{2}$. The bending energy, however, suppresses this soft mode by selecting a minimum in this class of isometric deformations: the base configuration having $\hat{\kappa}_{y}=\hat{c}_{y}$ and $\hat{\kappa}_{xy}=\hat{c}_{xy}$.

\begin{figure}[t]
\centering
\includegraphics[width=.9\textwidth]{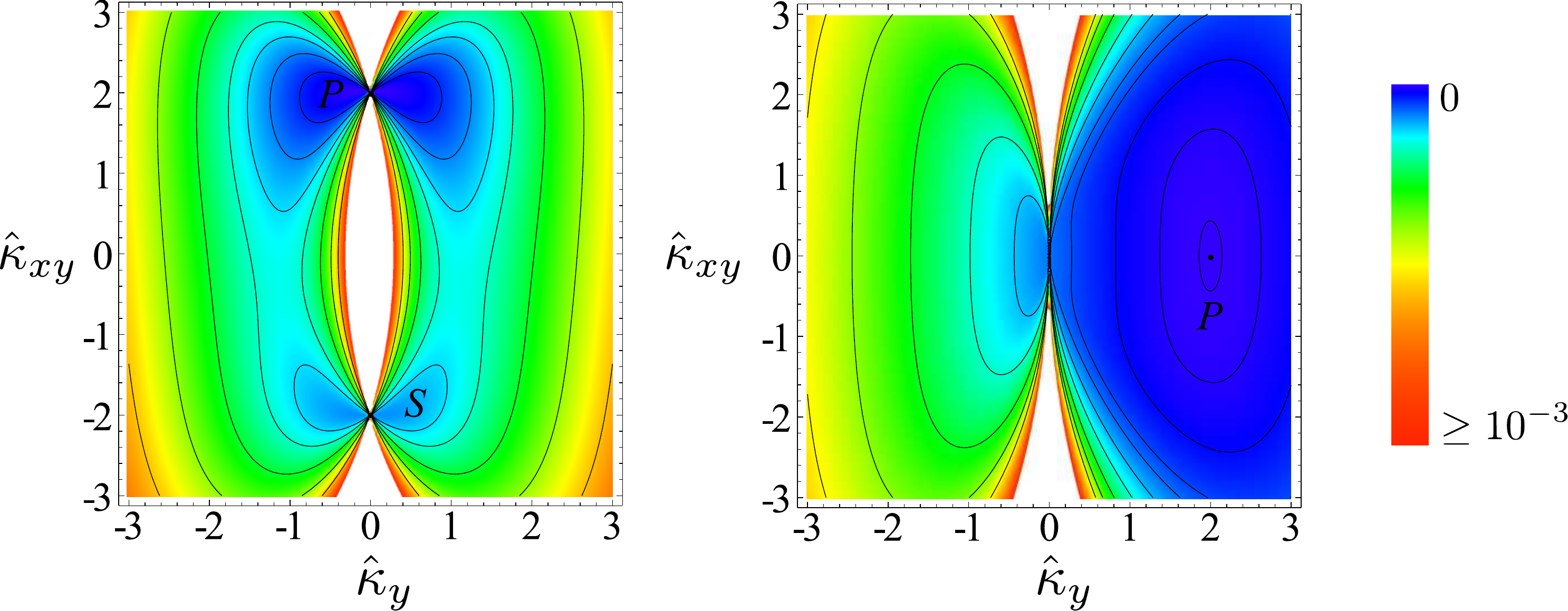}
\caption{\label{fig:contour1}Contour plot of the total elastic energy per unit length $\hat{U}$ from \eqref{eq:fs} and \eqref{eq:fb} in the plane $(\hat{\kappa}_{y},\,\hat{\kappa}_{xy})$. The labels {\em P} and {\em S} indicate the primary and secondary minima. On the left the case $\hat{c}_{x}=\hat{c}_{y}=0$ and $\hat{c}_{xy}=2$. The strip has a secondary minimum at $\hat{\kappa}_{xy}\approx -\hat{c}_{xy}$. On the right, $\hat{c}_{x}=\hat{c}_{xy}=0$ and $\hat{c}_{y}=2$. The strip has a single equilibrium configuration.} The other relevant parameters are $\beta=10$, $\rho=1$, $\nu=1/3$, $\hat{t}=10^{-2}$.
\end{figure}

A special but not necessarily isometric solution with constant curvature can be found by taking the limit $\hat{\kappa}_{y}\rightarrow 0$ of equation (\ref{eq:orthotropic-solution}a) which gives:
\begin{equation}
\hat{\kappa}_{x} = \hat{c}_{x}+\nu \hat{c}_{y}\,.	
\end{equation}
The corresponding longitudinal stress is given then by:
\begin{equation}\label{eq:longitudinal-stress}
\hat{N}_{y} = \frac{\hat{t}}{6}\,\beta\,(3\hat{x}^{2}-1)[(\hat{c}_{x}\hat{c}_{y}-\hat{c}_{xy}^{2})+\hat{\kappa}_{xy}^{2}]\,,
\end{equation}
using which we can calculate the total elastic energy as:
\begin{equation}\label{eq:zero-kappa-y}
\hat{U} 
= \beta \left\{ \frac{1}{45}\,[(\hat{c}_{x}\hat{c}_{y}-\hat{c}_{xy}^{2})+\hat{\kappa}_{xy}^{2}]^{2}
+ \frac{\hat{t}^{2}}{12}\,\left[\hat{c}_{y}^{2}+\frac{\rho}{\beta}\,(\hat{\kappa}_{xy}-\hat{c}_{xy})^{2}\right]  
\right\}\,.
\end{equation}
This class of deformations belongs to the set of isometries only when the first term in the left-hand side of equation \eqref{eq:zero-kappa-y} vanishes, namely when:
\begin{equation}\label{eq:twist-negative-curvature}
\hat{\kappa}_{xy}^{2} = \hat{c}_{xy}^{2}-\hat{c}_{x}\hat{c}_{y} = -\hat{K}_{0}\,.
\end{equation}
Clearly this is possible only when the initial Gaussian curvature $\hat{K}_{0}$ is negative or zero. In the latter case the strip has only one minimum corresponding to the base configuration (see right of Figure \ref{fig:contour1}). 

However, a secondary equilibrium configuration of constant curvature is obtained if $\hat{c}_{x}=\hat{c}_{y}=0$ and $\hat{c}_{xy}\ne0$, when Eq. \eqref{eq:zero-kappa-y} reads:
\begin{equation}\label{eq:pure-twist}
\hat{U} 
= \beta\left\{ \frac{1}{45}\,(\hat{\kappa}_{xy}^{2}-\hat{c}_{xy}^{2})^{2}
+ \frac{\hat{t}^{2}}{12}\frac{\rho}{\beta}\,(\hat{\kappa}_{xy}-\hat{c}_{xy})^{2})
\right\}\,.
\end{equation}
Then, in addition to the base configuration $\hat{\kappa}_{xy}=\hat{c}_{xy}$, the energy has a secondary minimum corresponding to the twisting curvature:
\begin{equation}\label{eq:inverse-twist}
\hat{\kappa}_{xy} = -\frac{1}{2}\left\{\hat{c}_{xy}+\sqrt{\hat{c}_{xy}^{2}-\frac{15}{2}\,(\rho/\beta)\,\hat{t}^{2}}\right\}\,,
\end{equation}
which is approximately $\hat{\kappa}_{xy}\approx -\hat{c}_{xy}$ (see Figure \ref{fig:contour1}, left). The stability of the secondary minimum depends on $\rho$, $\beta$ as well as the Poisson ratio $\nu$. To assess this, we calculate the components of the Hessian matrix:
\begin{gather*}
\frac{\partial^{2}\hat{U}}{\partial \hat{\kappa}_{y}^{2}} = \beta\left\{\frac{\hat{t}^{2}}{6}+\frac{4}{45}\nu\,(\hat{\kappa}_{xy}^{2}-\hat{c}_{xy}^{2})+\frac{16}{945}\frac{(\nu^{2}-\beta)}{\hat{t}^{2}}(\hat{\kappa}_{xy}^{2}-\hat{c}_{xy}^{2})^{2}\right\}\,,\\[10pt]
\frac{\partial^{2}\hat{U}}{\partial\hat{\kappa}_{y}\partial\hat{\kappa}_{xy}} = 0\,,\qquad\qquad
\frac{\partial^{2}\hat{U}}{\partial \hat{\kappa}_{xy}^{2}} = -\frac{4}{45}\beta\,(\hat{c}_{xy}^{2}-3\kappa_{xy}^{2})+\frac{\hat{t}^{2}}{6}\rho\,,
\end{gather*}
where $\hat{\kappa}_{xy}$ is given by Eq. \eqref{eq:inverse-twist}. In order for the secondary minimum to be stable, the eigenvalues of the Hessian matrix must be positive. Upon expanding $\partial^{2}\hat{U}/\partial\hat{\kappa}_{y}^{2}$ at the second order in $\hat{t}$, this condition reduces to the following inequality: 
\[
\beta -2\nu\rho-\frac{10}{7}\left(1-\frac{\nu^{2}}{\beta}\right)\rho^{2}>0\,.
\]
This implies the following requirment on $\rho$:
\begin{equation}\label{eq:twist-stability}
\rho < \frac{\beta}{\nu+\sqrt{\frac{10\beta-3\nu^{2}}{7}}}\,.
\end{equation}
In summary, strips having zero spontaneous principal curvatures $\hat{c}_{x}=\hat{c}_{y}=0$ and non-zero spontaneous twist $\hat{c}_{xy}$, admit a secondary equilibrium configuration where the principal curvatures vanish, i.e. $\hat{\kappa}_{x}=\hat{\kappa}_{y}=0$ while the twisting curvature is reversed $\hat{\kappa}_{xy}\approx-\hat{c}_{xy}$. The secondary minimum is stable for low enough shear modulus, a scenario previously discussed by Seffen (2007) for the case of an elliptical plate endowed with pure twist. The most important difference between the two cases is that here the strip has zero bending moment along the lateral edges unlike the plate investigated by Seffen which can support a finite bending moment at the boundary. As a consequence the transverse curvature $\hat{\kappa}_{x}$, which is a free parameter in uniform curvature models, is here adjusted in order to maintain the boundary moment-free and this severely restricts the space of possible uniform curvature configurations. 

The previous situation is an example of bistability in a strip with a negative spontaneous Gaussian curvature $\hat{K}_{0}=-\hat{c}_{xy}^{2}$, where  the principal curvatures $\hat{c}_{x}$ and $\hat{c}_{y}$ vanish identically while the surface is purely twisted. However when $\hat{c}_{x}\hat{c}_{y}\ne 0$ and $\hat{K}_{0}<0$, the strip also exhibits a secondary equilibrium configuration, albeit with a variable transverse curvature $\hat{\kappa}_{x}$   discussed in the next section. 

\subsection{\label{sec:3b}Non-uniform curvature configurations}

\begin{figure}[t]
\centering
\includegraphics[width=.9\textwidth]{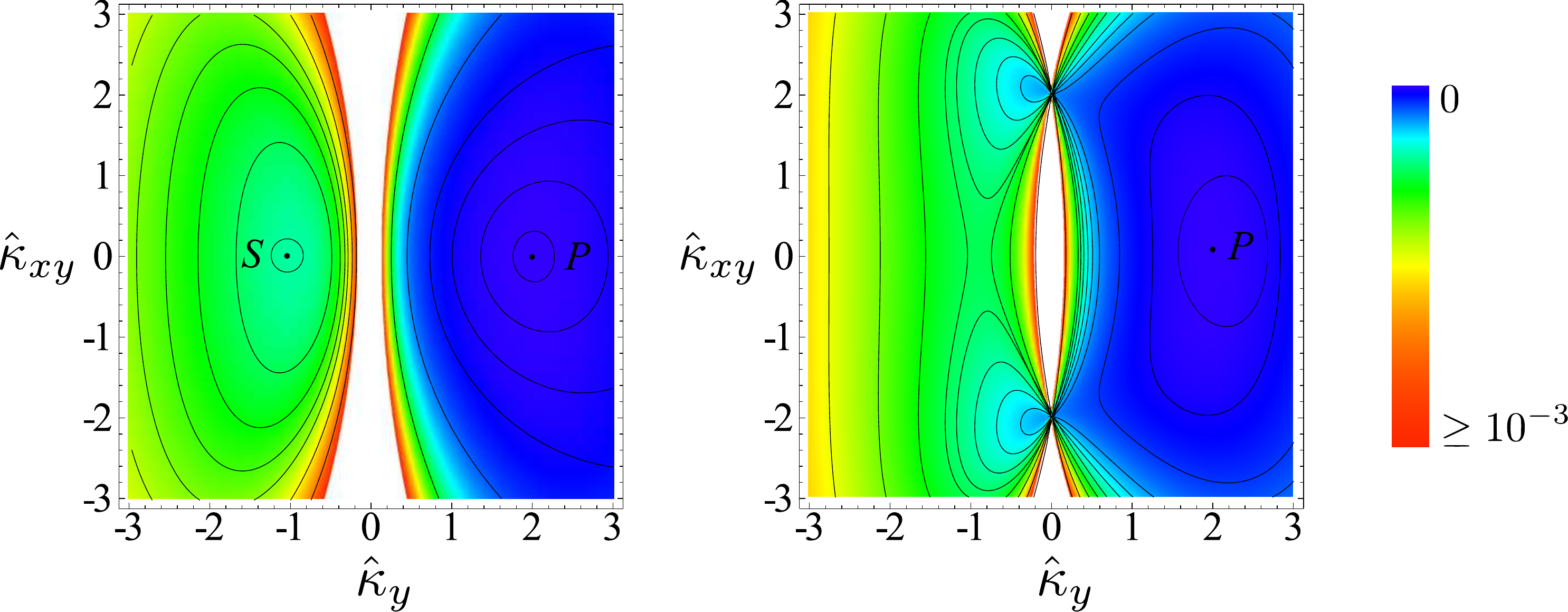}
\caption{\label{fig:contour2}Contour plot of the total elastic energy per unit length $\hat{U}$ from \eqref{eq:fs} and \eqref{eq:fb} in the plane $(\hat{\kappa}_{y},\,\hat{\kappa}_{xy})$. The labels {\em P} and {\em S} indicate the primary and secondary minima. On the left the case $\hat{c}_{x}=\hat{c}_{y}=2$ and $\hat{c}_{xy}=0$. In addition to the base configuration, the strip has a second equilibrium at $\hat{\kappa}_{y}\approx -(\hat{c}_{x}\hat{c}_{y})^{1/2}$. On the right, $\hat{c}_{x}=-2$, $\hat{c}_{y}=2$ and $\hat{c}_{xy}=0$. In this case the strip has only one equilibrium configuration. Note that the region $\hat{\kappa}_{y}<0$ that in the previous example was separated form the region $\hat{\kappa}_{y}>0$ by an energy barrier of height $\hat{U}_{0}\sim \hat{K}_{0}^{2}$ is now connected by two ``passes'' located at $\hat{\kappa}_{xy}^{2}=\pm(\hat{c}_{x}\hat{c}_{y})^{1/2}$. The other relevant parameters are $\beta=10$, $\rho=1$, $\nu=1/3$, $\hat{t}=10^{-2}$.}
\end{figure}

\begin{figure}
\centering
\includegraphics[width=.9\textwidth]{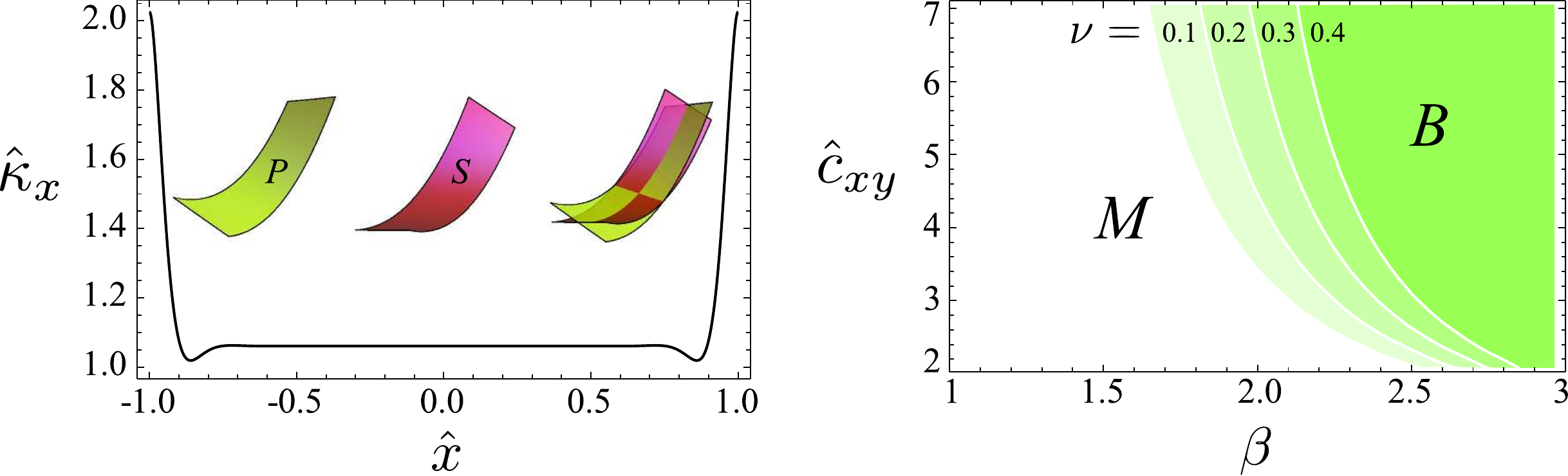}
\caption{\label{fig:boundary-layer}(Left) The dimensionless transverse curvature $\hat{\kappa}_{x}$ associated with the secondary equilibrium configuration of a strip with spontaneous positive Gaussian curvature as a function of the dimensionless distance $\hat{x}$ from the centerline. The relevant parameters are $\hat{c}_{x}=2$, $\hat{c}_{y}=1$, $\hat{c}_{xy}=0$, $\beta=10$, $\rho=1$ and $\hat{t}=10^{-2}$. The transverse curvature is constant and equal to $\kappa_{0}$ with exception for a boundary layer. In the inset a schematic representation of the primary ($P$) and secondary ($S$) equilibria. (Right) Monostability/bistability ({\em M}/{\em B}) phase diagram in the plane $(\beta,\,\hat{c}_{xy})$ for a strip of negative Gaussian curvature and different values of the Poisson ratio $\nu$. The spontaneous principal curvatures are $\hat{c}_{x}=2$ and $\hat{c}_{y}=1$. The other relevant parameters are $\rho=1$ and $\hat{t}=10^{-2}$.}
\end{figure}

When the constraint of constant curvature is lifted, new minima appear in the energy landscape. Figure \ref{fig:contour2} shows a contour plot of the total elastic energy in the plane $(\hat{\kappa}_{x},\,\hat{\kappa}_{xy})$ for the two cases $\hat{c}_{x}\hat{c}_{y}>0$ (left), $\hat{c}_{x}\hat{c}_{y}<0$ (right) and $\hat{c}_{xy} = 0$ in both case. For positive spontaneous Gaussian curvatures the energy minima are those described in the previous section when $\hat{\kappa}_{xy}=0$. 
The transverse curvature $\hat{\kappa}_{x}$ associated with the secondary minimum at $\hat{\kappa}_{y}<0$ is mostly constant across the width of the strip and equal to
\[
\hat{\kappa}_{0} = \frac{(\hat{c}_{x}\hat{c}_{y}-\hat{c}_{xy}^{2})+\hat{\kappa}_{xy}^{2}}{\hat{\kappa}_{y}}\,,
\]
with exception for a boundary layer (Figure \ref{fig:boundary-layer}). This phenomenon,  first noted by Lamb (1891), is due to the rapid build-up of the bending moment $M_{x}$ from zero at the edges to a non-zero value that inevitably develops in any configuration other than the base state (see Mansfield 1989 for a detailed explanation). An estimate of the size of the boundary layer is given by $1/k\approx(t/|\kappa_{y}|)^{\frac{1}{2}}$; however it vanishes in a strip of lenticular cross section whose thickness (hence the bending moment) smoothly tapers and vanishes at the edges.

\begin{figure}[t]
\centering
\includegraphics[width=0.9\textwidth]{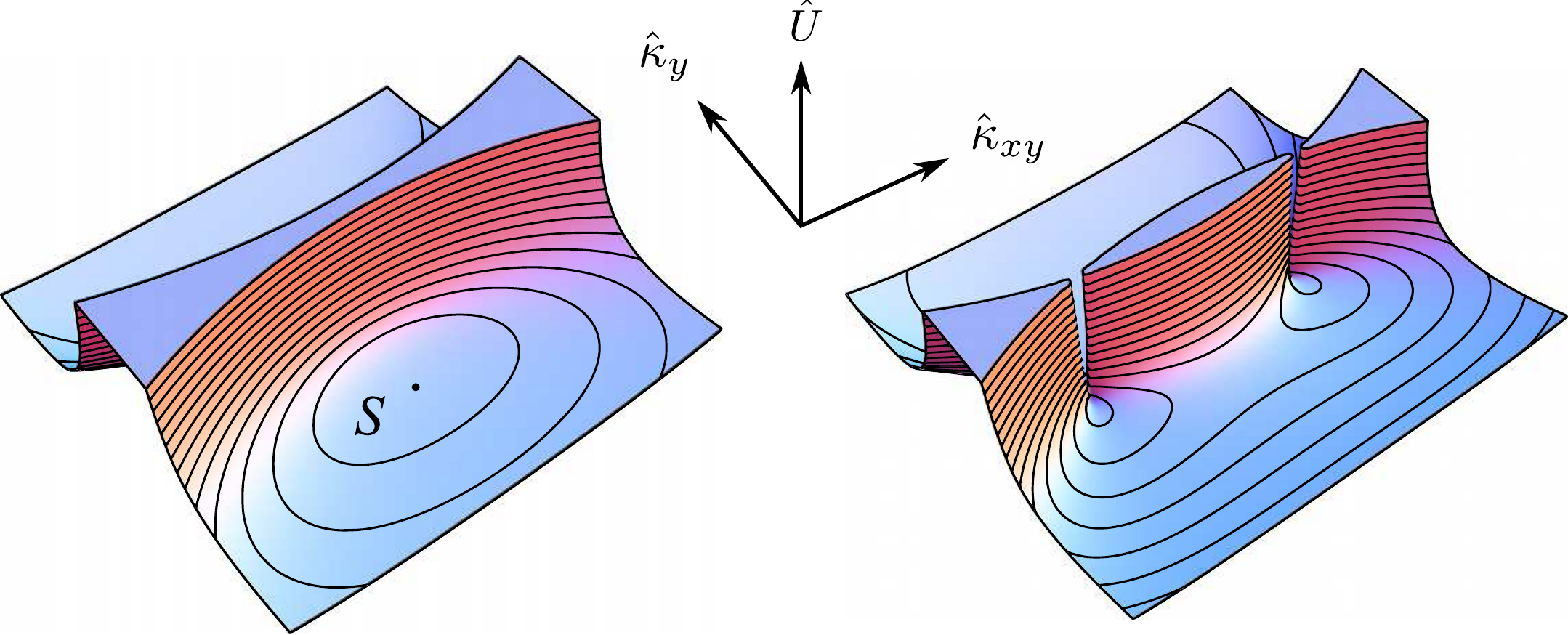}
\caption{\label{fig:landscapes}The elastic energy given by \eqref{eq:fs} and \eqref{eq:fb} as function of $\hat{\kappa}_{y}$ and $\hat{\kappa}_{xy}$ for a strip of spontaneous positive Gaussian curvature (on the left, $\hat{c}_{x}=\hat{c}_{y}=1$) and a strip of spontaneous negative Gaussian curvature (on right, $\hat{c}_{x}=-\hat{c}_{y}=0$). In the former, the $\hat{\kappa}_{y}<0$ portion of the energy landscape is separated from the $\hat{\kappa}_{y}>0$ region by a large energy barrier of height $\hat{U}(\hat{\kappa}_{y}=0)\sim \hat{K}_{0}^{2}$ . This implies the existence of a secondary minimum (labelled as {\em S}). In the case of strips of negative spontaneous Gaussian curvature, on the other hand, the energy barrier has two passes at $\hat{\kappa}_{y}=0$ and $\hat{\kappa}_{xy}=\pm(-\hat{K}_{0})^{1/2}$ and the secondary minimum is replaced by a saddle point. The flat plateau in the energy barriers is a graphical artifact due to the limited range of $\hat{U}$ shown.}	
\end{figure}

On on the other hand, for negative Gaussian curvature $\hat{c}_{x}\hat{c}_{y}<0$ and $\hat{c}_{xy}=0$, the secondary minimum becomes a saddle point and the elastic energy has only one minimum corresponding to the base configuration. It is interesting to notice, in this latter case, that the regions of positive and negative $\hat{\kappa}_{y}$ of the energy landscape are separated by a barrier of height $\hat{U}(\hat{\kappa}_{y}=0)\sim \hat{K}_{0}^{2}$ that is large everywhere with exception for two ``passes'' at $\hat{\kappa}_{xy}=\pm(-\hat{c}_{x}\hat{c}_{y})^{1/2}$ (see Figure \ref{fig:landscapes}). The existence of these passes along the energy barrier at $\hat{\kappa}_{y}=0$ is the reason for the non-existence of a secondary equilibrium configuration in strips with spontaneous negative Gaussian curvature. As explained in the previous section and summarized in equation \eqref{eq:twist-negative-curvature}, the stretching energy associated with the barrier at $\hat{\kappa}_{y}=0$ scales like $\hat{U}_{s}\sim(\hat{K}_{0}+\hat{\kappa}_{xy}^{2})^{2}$ and, in presence of spontaneous {\em negative} Gaussian curvature, it can be relieved through twist. This leads to the formation of passes at $\hat{\kappa}_{xy}=\pm(-\hat{K}_{0})^{1/2}$ shown in Figure \ref{fig:contour2} and \ref{fig:landscapes} which allow the transverse curvature $\hat{\kappa}_{y}$ to be switched from positive to negative (and vice versa) {\em isometrically}. Positive spontaneous Gaussian curvatures, on the other hand, cannot be accommodated through twist and thus there are no passages across the energy barrier at $\hat{\kappa}_{y}=0$; consequently there is a secondary minimum. 

\begin{figure}[h!]
\centering
\includegraphics[scale=0.4]{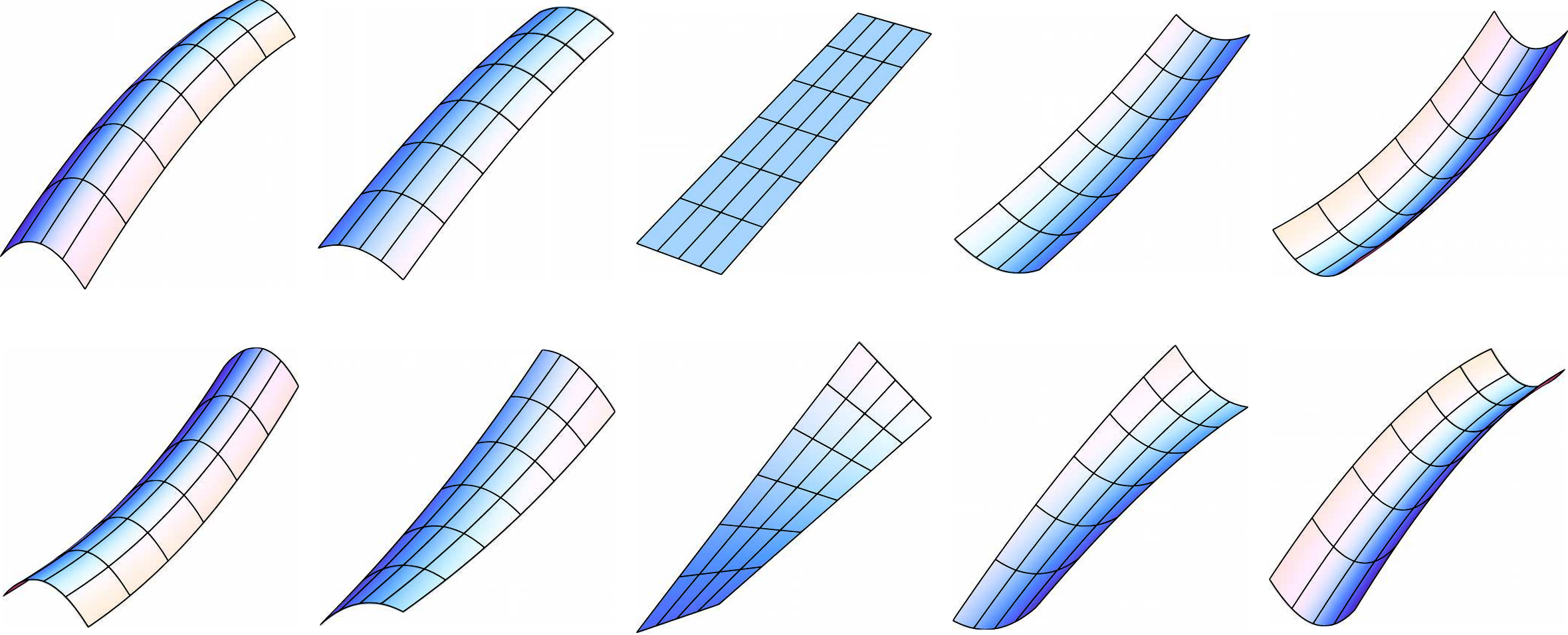}
\caption{\label{fig:frames}Sequence of deformations of a strip of spontaneous positive (top) and negative (bottom) Gaussian curvature associated with curvature reversal. While the former sequence requires the strip to become temporarily flat and thus produces large strains, the latter can be performed isometrically at zero stretching energy cost. The existence of this low energy deformation in strips of spontaneous negative curvature is responsible for the formation of the passes shown in Figure \ref{fig:landscapes}, hence the lack of a secondary minimum.}
\end{figure} 

To get an intuitive sense of these  passes in the energy landscape, imagine a strip of  positive spontaneous Gaussian curvature (upper left frame of Figure \ref{fig:frames}). To evert the strip, one choice is to flatten the strip and then reverse its curvature. Since the strip is naturally curved, the intermediate flat conformation will suffer a large strain and this corresponds to the large energy barrier shown in Figure \ref{fig:landscapes}. On the other hand, if the strip is naturally shaped like a saddle (bottom left frame of Figure \ref{fig:frames}), turning the strip inside-out can be achieved by pulling apart two opposite corners until the strip is fully extended and maximally twisted, and then bend it into the new configuration. The latter deformation can be performed without stretching the surface and thus with no energy cost other than bending, in contrast with the case of a strip with positive $\hat{K}_{0}$. Similar deformation pathways have been analyzed by Fernandes {\em et al}. (2010) as possible strategies to achieve  shape control of bistable composite plates using embedded actuators. 

\begin{figure}[t]
\centering
\includegraphics[width=.9\textwidth]{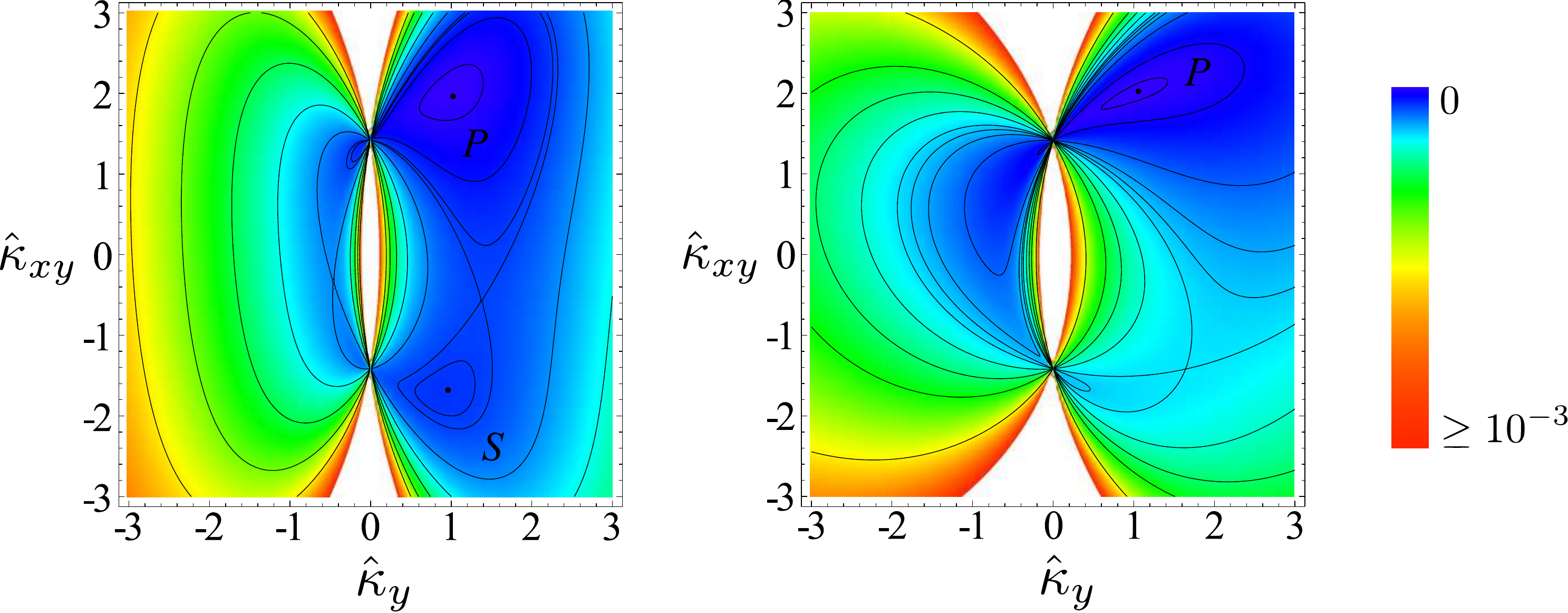}
\caption{\label{fig:contour3}Contour plot of the total elastic energy per unit length $\hat{U}$ from \eqref{eq:fs} and \eqref{eq:fb} in the plane  $(\hat{\kappa}_{y},\,\hat{\kappa}_{xy})$. The labels {\em P} and {\em S} indicate the primary and secondary minima. In both plots the spontaneous curvatures are set $\hat{c}_{x}=2$, $\hat{c}_{y}=1$ and $\hat{c}_{xy}=2$. On the left, the case $\beta=10$, in addition to the base configuration, the strip as a secondary minimum in proximity of the point $(\hat{\kappa}_{y},\,\hat{\kappa}_{xy})=(\hat{c}_{y},\,-\hat{c}_{xy})$. Such a minimum disappear when $\beta=1$ (right). The other relevant parameters are $\rho=1$, $\nu=1/3$, $\hat{t}=10^{-2}$.}
\end{figure}

However, when strips with negative spontaneous Gaussian curvature $\hat{K}_{0}<0$ have a non-zero spontaneous twist $\hat{c}_{xy}$, they admit bistable configurations. In addition to the configuration with constant curvature described in the previous section in the special case $\hat{c}_{x}=\hat{c}_{y}=0$ and $\hat{c}_{xy}\ne 0$, another configuration can be obtained by setting $\hat{c}_{x}\hat{c}_{y}\ne 0$ with $\hat{K}_{0}<0$. Figure \ref{fig:contour3} shows a contour plot of the total elastic energy \eqref{eq:fs} and \eqref{eq:fb} of a strip with $\hat{c}_{x}=2$, $\hat{c}_{y}=1$ and $\hat{c}_{xy}=2$, thus $\hat{K}_{0}=-2$. The plot on the left (in which $\beta=10$) shows a secondary minimum in the proximity of the point $(\hat{\kappa}_{y},\,\hat{\kappa}_{xy})=(\hat{c}_{y},\,-\hat{c}_{xy})$. When anisotropy parameter $\beta$ is decreased, however, the secondary minimum disappears and the energy landscape is again characterized by a unique minimum corresponding to the base configuration. 

Such a configuration has energy $\hat{U}\approx (\rho/\beta)\,\hat{c}_{xy}^{2}$, and has a minimum for large values of $\beta/\rho$. One might expect that the value of $\beta/\rho$ necessary to have a secondary minimum of this kind would increase monotonically with $\hat{c}_{xy}$; however this is incorrect. In Figure \ref{fig:boundary-layer} we show the ``phase diagram'' of the monostability/bistability region for a strip with spontaneous curvatures $\hat{c}_{x}=2$ and $\hat{c}_{y}=1$ in the plane $(\beta/\rho,\,\hat{c}_{xy})$. Upon fixing the spontaneous twist $\hat{c}_{xy}$, the bistability region can be accessed by increasing $\beta/\rho$ as expected. However, for fixed $\beta/\rho$, the bistable region is accessed by increasing the spontaneous twist. The origin of this behavior is related with the height of the energy barrier at $\hat{\kappa}_{y}=0$ separating the secondary minimum from the basin of attraction of the base configuration at $\hat{\kappa}_{y}<0$. From equation \eqref{eq:zero-kappa-y} we see that the height of such a barrier  $\hat{U}\approx\beta \hat{K}_{0}^{2}$. For fixed $\beta/\rho$, decreasing $\hat{c}_{xy}$ (or equivalently $|\hat{K}_{0}|$) will lower the height of the barrier until the basin of attraction associated with the portion where $\hat{\kappa}_{y}<0$  of the base configuration will merge with that of the secondary minimum. For values of $\hat{c}_{xy}$ smaller than this critical value, the energy landscape is then characterized by a unique minimum (i.e. the base configuration) surrounded by a large {\em C}-shaped basin of attraction that starts from the base configuration and extends in the region $\hat{\kappa}_{y}<0$ penetrating the barrier at $\hat{\kappa}_{y}=0$ through the passes located at $\hat{\kappa}_{xy}=\pm (-\hat{K}_{0})^{1/2}$ (see right of Figure \ref{fig:contour3}).  Once again, the presence of two regions across the barrier at $\hat{\kappa}_{y}=0$ where the stretching energy drops, plays a crucial role in assuring the existence of bistable configurations.

\begin{figure}
\centering
\includegraphics[width=1.\textwidth]{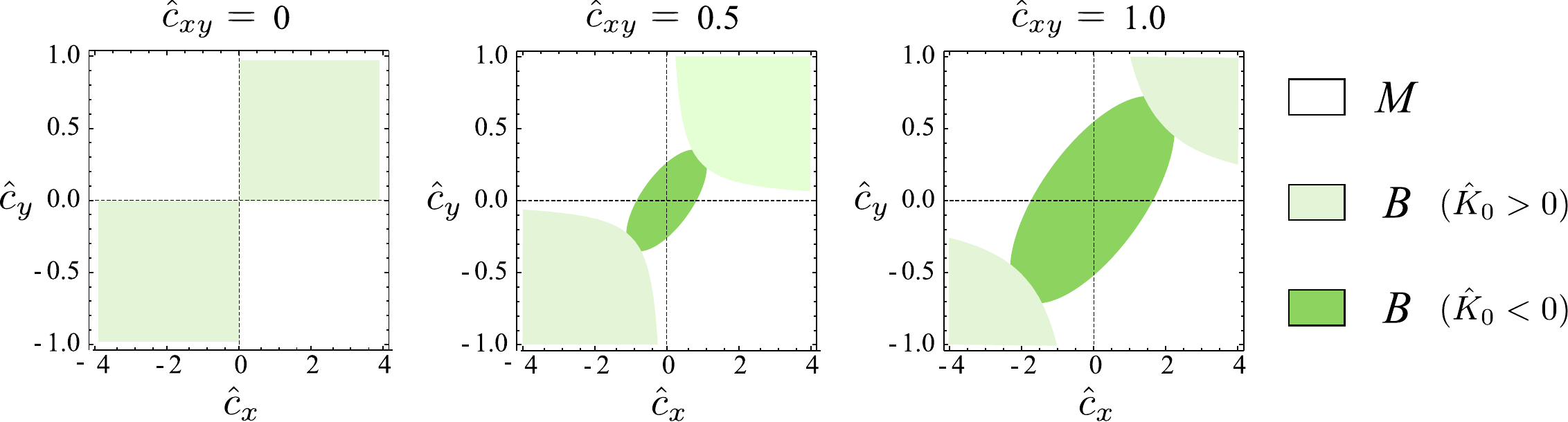}
\caption{\label{fig:phase-diagram1}Phase diagram of the monostability ({\em M}, in white) and bistability region ({\em B}, shaded) in the plane $(\hat{c}_{x},\,\hat{c}_{y})$ for strips of various spontaneous twist $\hat{c}_{xy}$. \revision{Different shades of green correspond to positve (light) and negative (dark) spontaneous Gaussian curvature}. For zero spontaneous twist, bistable configuration exist only in the region of positive Gaussian curvature. For $\hat{c}_{xy}\ne 0$, on the other hand, bistability also occurs in a range of negative Gaussian curvature. The relevant parameters are $\beta=10$, $\rho=1$, $\nu=1/3$ and $\hat{t}=10^{-2}$.}
\end{figure}

In summary, orthotropic strips exhibit various form of bistability depending to their spontaneous curvature and elastic moduli. The simplest bistable configurations have constant curvature throughout the width of the strip and correspond to isometries of the base configuration. Non-isometric secondary equilibria have, on the other hand, variable transverse curvature $\hat{\kappa}_{x}$; this is approximately constant and equal to $\hat{\kappa}_{0}=(\hat{K}_{0}+\hat{\kappa}_{xy}^{2})/\hat{\kappa}_{y}$ in the bulk of the strip, but abruptly jumps to $\hat{\kappa}=\hat{c}_{x}+\nu(\hat{c}_{y}-\hat{\kappa}_{x})$ near the edges to relieve the bending moment that forms in the bulk when the strip is in a configuration different from the base one. For strips with positive spontaneous Gaussian curvature, in particular, the secondary equilibrium configuration has transverse curvature of opposite sign relative to the base configuration. Strips with negative spontaneous Gaussian curvature, on the other hand, do not possess a secondary equilibrium configuration unless they have sufficient spontaneous twist $\hat{c}_{xy}$. The existence of bistability in this case also relies on the value of the ratio $\beta/\rho$ between the Young's modulus along the longitudinal direction and the shear modulus. Figure \ref{fig:phase-diagram1} shows a phase-diagram of the monostability/bistability regions in in the plane $(\hat{c}_{x},\,\hat{c}_{y})$ for strips of various spontaneous twist $\hat{c}_{xy}$.

\subsection{\label{sec:3c}Tristable configurations}

As we mentioned in the introduction of Section \ref{sec:3}, the requirement for the matrices ${\bf A}$ and ${\bf D}$ to be positive definite translates into the conditions $\beta>\nu^{2}$ and $\rho>0$. In this section we show that when the limit $\beta\rightarrow\nu^{2}$ is approached, a new stable minimum appears in the energy landscape of the configurations allowing for {\em tristability}. The presence of three minima in the reduced quasi-one-dimensional geometry of elastic strips complements the appearance of such a landscape for corrugated strips, where tristability arises as the consequence of the coupling between internal prestresses created while imprinting the corrugation and non-linear geometrical changes during deformation (Norman {\em et al}. 2008). 

\begin{figure}[t]
\centering
\includegraphics[width=.9\textwidth]{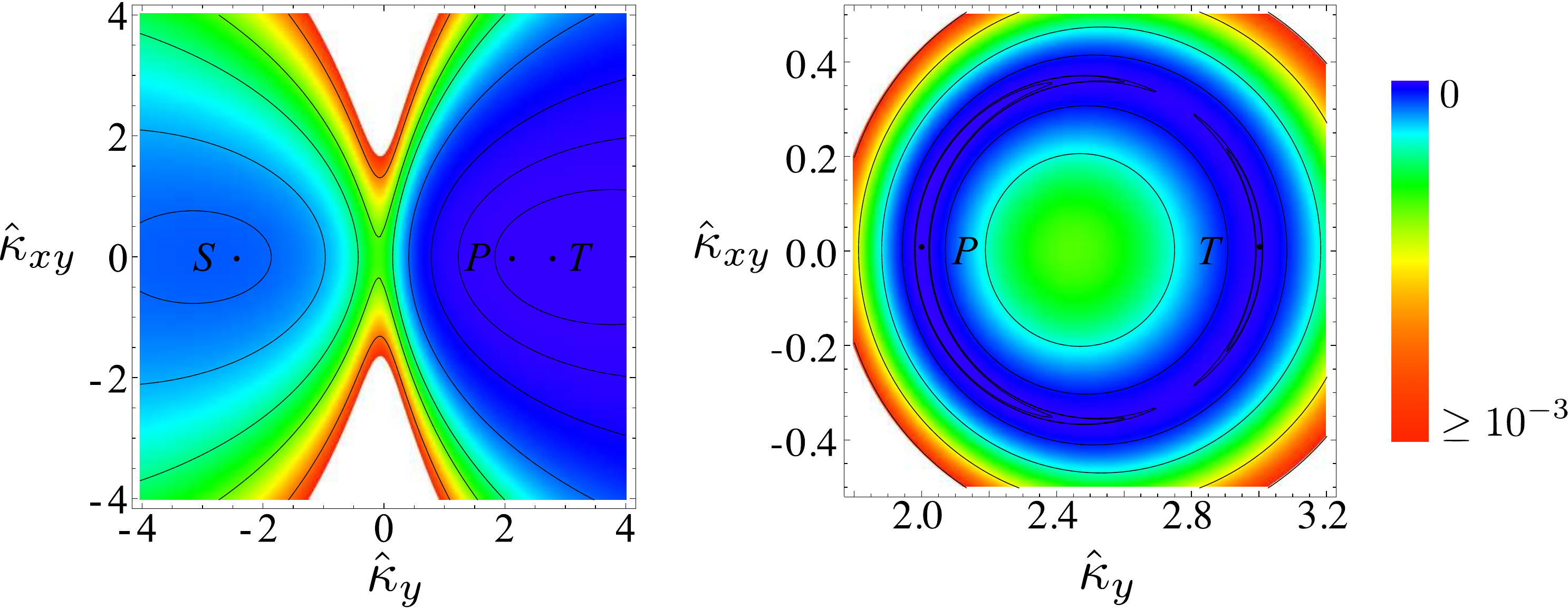}	
\caption{\label{fig:contour4}Contour plot of the total elastic energy per unit length $\hat{U}$ from \eqref{eq:fs} and \eqref{eq:fb} in the plane  $(\hat{\kappa}_{y},\,\hat{\kappa}_{xy})$ of a strip with $\beta=\nu^{2}+10^{-5}$ and spontaneous curvatures $\hat{c}_{x}=1$, $\hat{c}_{y}=2$ and $\hat{c}_{xy}=0$. The primary, secondary and tertiary minima are indicated with the letters $P$, $S$ and $T$ respectively. The plot on the right shows a magnification of the region $\hat{\kappa}_{y}\in[1.8,\,3.2]$ where the primary and tertiary minimum are located. The other relevant parameters are $\rho=1$, $\nu=1/3$, $\hat{t}=5\times 10^{-4}$.}
\end{figure}

Letting $\beta=\nu^{2}+\epsilon$ and expanding equations \eqref{eq:fs} and \eqref{eq:fb} in powers of $\epsilon$ gives:
\begin{multline}\label{eq:tristable-energy}
\hat{U} 
= \nu^{2}\bigg\{
\frac{1}{45}\,\left[(\hat{\kappa}_{y}-\hat{c}_{y})(\nu\hat{\kappa}_{y}-\hat{c}_{x})+\hat{\kappa}_{xy}^{2}-\hat{c}_{xy}^{2}\right]^{2}\\
+\frac{1}{12}\,\hat{t}^{2}\left[(\hat{\kappa}_{y}-\hat{c}_{y})^{2}+\frac{\rho}{\nu^{2}}\,(\hat{\kappa}_{xy}^{2}-\hat{c}_{xy}^{2})\right]
\bigg\}+o(\epsilon)\,.
\end{multline}
Then, with $\hat{\kappa}_{xy}=\hat{c}_{xy}$, it is easy to prove that $\partial \hat{U}/\partial \hat{\kappa}_{y}=0$   when $\hat{\kappa}_{y}=\hat{c}_{y}$ and also when:
\begin{equation}\label{eq:kappa-pm}
\hat{\kappa}_{y}^{\pm} = \frac{1}{4}\left\{\hat{c}_{y}+\frac{1}{\nu}\left[3\hat{c}_{x}\pm\sqrt{(\hat{c}_{x}-\nu \hat{c}_{y})^{2}-30\,\hat{t}^{2}}\,\right]\right\}\,.
\end{equation}
Thus, in addition to the base configuration and the secondary minimum at $\hat{\kappa}_{y}<0$ discussed in section (\ref{sec:3b}), the energy has a third minimum at $\hat{\kappa}_{y}=\hat{\kappa}_{y}^{+}$ while $\hat{\kappa}_{y}^{-}$ is the location of the maximum separating the third minimum from the base configuration. For relatively thin shells, $\hat{t}\ll1$ so that one finds simply $\hat{\kappa}_{y}^{+}\approx \hat{c}_{x}/\nu$ and $\hat{\kappa}_{y}^{-}\approx(\hat{c}_{y}+\hat{c}_{x}/\nu)/2$. The energy of the third minimum is given approximately by:
\begin{equation}
\hat{U}(\hat{\kappa}_{y}^{+})\approx \frac{1}{12}\,\hat{t}^{2}\nu^{2}\left(\hat{c}_{y}-\frac{\hat{c}_{x}}{\nu}\right)^{2}
\end{equation}
Figure \ref{fig:contour4} shows a contour plot of the total elastic energy per unit length $\hat{U}$ from \eqref{eq:fs} and \eqref{eq:fb} for $\hat{c}_{x}=1$, $\hat{c}_{y}=2$, $\hat{c}_{xy}=0$ and $\beta=\nu^{2}+10^{-2}$, with $\nu=1/3$ and $\rho=1$. The primary, secondary and tertiary minima are indicated with the letters $P$, $S$ and $T$ respectively. 

In the limit $\beta\rightarrow\nu^{2}$, tristability has been noted by Vidoli \& Maurini (2009) in the case of plates with free boundaries and  uniform curvature, although the physical origins of this phenomenon were not discussed. In our simple setting,  consistent with the arguments in section (\ref{sec:3a}), there is a continuous set of configurations of zero stretching energy corresponding to the isometries of the base state. In the plane $(\hat{\kappa}_{y},\,\hat{\kappa}_{xy})$ this set describes an ellipse whose intersections with the $\hat{\kappa}_{y}$ axis (i.e. the untwisted configurations) are given by $\hat{\kappa}_{y}=\hat{c}_{y}$ and $\hat{\kappa}_{y}=\hat{c}_{x}/\nu$. The small but finite bending stiffness along the principal directions prevents this from being a true zero energy mode and instead leads to the existence of a soft mode by raising the energy of the minimum at $\hat{\kappa}_{y}=\hat{c}_{x}/\nu$; for $\beta\gg\nu^{2}$ (i.e. $D_{22}\gg D_{11}$) the minimum is completely suppressed. Taking the limit $\beta\rightarrow\nu^{2}$ corresponds to lowering the bending stiffness so that the energies of the two isometric configurations $\hat{\kappa}_{y}=\hat{c}_{y}$ and $\hat{\kappa}_{y}=\hat{c}_{x}/\nu$ are comparable again. This mechanism is completely general and independent of the specific geometry of the problem; any plate with shallow spontaneous curvature is amenable to a set of isometries and  multistability can be in principle obtained by setting $\det({\bf D})\sim 0$, which in the case of orthotropic plates corresponds in fact to $\beta\sim\nu^{2}$.

An interesting feature of this tristable energy landscape relies on the fact that the energy barrier separating the minima at $\hat{\kappa}_{y}=\hat{c}_{y}$ and $\hat{\kappa}_{y}=\hat{c}_{x}/\nu$ is generally much smaller than that located at $\hat{\kappa}_{y}=0$ separating the base configuration with the minimum at $\hat{\kappa}_{y}<0$. More precisely:
\begin{equation}
\hat{U}(\hat{\kappa}_{y}^{-}) \approx \frac{1}{720}\,(\hat{c}_{x}-\nu\hat{c}_{y})^{4}
\end{equation}
while $\hat{U}(0)\approx (1/45)\,(\nu \hat{K}_{0})^{2}$. Assuming $\hat{\kappa}_{xy}=\hat{c}_{xy}=0$, the ratio between the height of these two energy barriers can be approximated as:
\begin{equation}\label{eq:ratio}
\frac{\hat{U}(\hat{\kappa}_{y}^{-})}{\hat{U}(0)}
\approx\frac{1}{16}\left[\left(\frac{\hat{c}_{x}}{\nu \hat{c}_{y}}\right)+\left(\frac{\hat{c}_{x}}{\nu \hat{c}_{y}}\right)^{-1}-2\right]^{2}
=\left[\sinh\frac{1}{2}\log\left(\frac{\hat{c}_{x}}{\nu \hat{c}_{y}}\right)\right]^{4}\,,
\end{equation}
Assuming $\hat{c}_{x}<\hat{c}_{y}$,  this ratio is  $O(10^{-2})$. For $\hat{c}_{x}>\hat{c}_{y}$ the previous relation is no longer valid as the bending energy, that was neglected to derive \eqref{eq:ratio} becomes relevant; however, the order of magnitude of the ratio $\hat{U}(\hat{\kappa}_{y}^{-})/\hat{U}(0)$ remains always of order $O(10^{-2})$.

\begin{figure}[t]
\centering
\includegraphics[width=.9\textwidth]{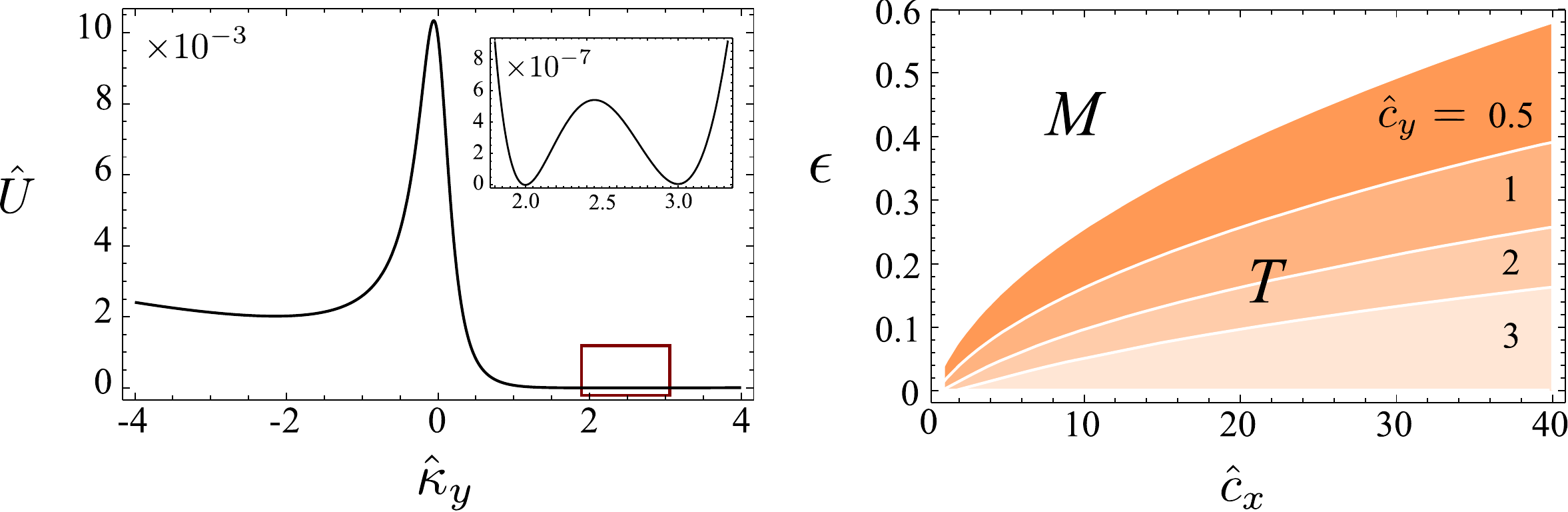}
\caption{\label{fig:phase-diagram2}(Left) Total elastic energy as a function of $\hat{\kappa}_{y}$ for a strip with $\beta-\nu^{2}=10^{-5}$ and spontaneous curvatures $\hat{c}_{x}=1$, $\hat{c}_{y}=2$ and $\hat{c}_{xy}=0$. The inset shows a magnification the boxed region $\hat{\kappa}_{y}\in [1.8,\,3.3]$. The height of the barrier separating the base configuration from the tertiary minimum located at $\kappa_{y}^{+}\approx c_{x}/\nu=3$ is four order of magnitude smaller than the barrier at $\hat{\kappa}_{y}=0$. (Right) phase-diagram in the plane $(\hat{c}_{x},\,\epsilon=\beta-\nu^{2})$ of the bistable ({\em M}, in white) and tristable ({\em T}, shaded) regions of a strip with zero spontaneous twist and various $\hat{c}_{y}$. The other relevant parameter are $\rho=1$ and $\hat{t}=10^{-2}$.}	
\end{figure}

The stability of the tertiary minimum at $\hat{\kappa}_{y}^{+}\approx \hat{c}_{x}/\nu$ depends, in general, on the value of the Poisson ratio and the spontaneous curvatures $\hat{c}_{x}$ and $\hat{c}_{y}$. Figure \ref{fig:phase-diagram2} shows the bistability/tristability regions as a function of $\epsilon=\beta-\nu^{2}$ and $\hat{c}_{x}$ for various values of $\hat{c}_{y}$. Increasing the transverse spontaneous $\hat{c}_{x}$ allows tristability to occur at larger values of $\epsilon$, thus further away from the limiting condition $\beta=\nu^{2}$.

\section{\label{sec:4}Strips with coupling between stretching and bending}

We now consider generally anisotropic strips, in which stretching and bending deformations are coupled through the ${\bf B}$ matrix, so that the total elastic energy density reads:
\begin{multline}
u = 
\frac{1}{2}\,\Big\{\alpha_{22}N_{y}^{2}+\delta_{11}(\kappa_{x}-c_{x})^{2}+2\delta_{12}(\kappa_{x}-c_{x})(\kappa_{y}-c_{y})\\+\delta_{22}(\kappa_{y}-c_{y})^{2}+\delta_{66}(\kappa_{xy}-c_{xy})^{2}\Big\}
\end{multline}
where the elastic moduli $\alpha_{22}$ and $\delta_{ij}$ are given by \eqref{eq:elastic-moduli} and $N_{y}$ and $\kappa_{x}$ have the general form given in equations \eqref{eq:anisotropic-stress} and \eqref{eq:anisotropic-curvature}. Because of the reduced dimensionality of this class of elastic objects, the coupling between stretching and bending is not expected to produce dramatic differences when compared to the case of orthotropic strips discussed in the previous section and indeed, for the majority of the choice of the elastic moduli $\alpha_{22}$ and $\delta_{ij}$, the behavior is qualitatively identical to that already described. 

Under some circumstance, however, the finer control of the material properties characteristic of anisotropic laminates might be taken advantage of to enhance the region of tristability. To understand this possibility, we first note that the requirement for the matrices $\boldsymbol{\alpha}$ and $\boldsymbol{\delta}$ to be positive-definite, translates here into the condition $\alpha_{22}>0$, $\delta_{66}>0$ and $\delta_{11}\delta_{22}>\delta_{12}^{2}$. If the latter condition, in particular, is only weakly satisfied, the strip will be tristable for some value of the spontaneous curvatures $c_{x}$ and $c_{y}$. To illustrate this point let us consider the case of an antisymmetric angle-ply laminates. The condition $\delta_{11}\delta_{22}>\delta_{12}^{2}$ translate into the following inequality:
\[
(B_{16}^{2}-A_{66}D_{11})(B_{26}^{2}-A_{66}D_{22})>(B_{16}B_{26}-A_{66}D_{12})^{2}\,.
\]
Regardless the specific values of $B_{ij}$ and $D_{ij}$, the left and right hand sides of this inequality will be comparable in any material where $B^{2}\gg A_{66}D$. In other words the limit $\det(\boldsymbol{\delta})\rightarrow 0$, that for orthotropic strips can be approached by taking $\beta \rightarrow \nu^{2}$, can here be obtained by increasing the strength of the coupling between stretching and bending. Thus we might expect angle-ply laminated strips with strong coupling between bending and stretching to be tristable for a broad range of spontaneous curvatures.

\section{\label{sec:5}Conclusions}

Despite its relatively long history, the theory of elastic mulitstability is still a field rich in challenging problems and open questions. Most of the theoretical work has been limited to the case of uniform Gaussian curvature deformations. Here, we lift the assumption of uniform curvature variations and inextensibility, and find a variety of morphing and multistable scenarios  in shells that have both anisotropy and spontaneous curvature. Our results complement and extend those previously reported by Galletly \& Guest (2004a, 2004b), Guest \& Pellegrino (2006), Seffen (2007) and Vidoli \& Maurini (2009).  In particular we emphasize the fundamental role of the intrinsic geometry of the base configuration embodied in the spontaneous Gaussian curvature of the strip. Both strips of positive and negative spontaneous Gaussian curvature admits bistable configurations, but while strips of positive spontaneous curvature are {\em always} bistable, regardless the values of the elastic moduli, the occurrence of bistability in strips of spontaneous negative curvature depends crucially on the presence of spontaneous twist as well as the relative stiffness of the strip under tensile and shear deformations. This fundamental difference stems from the fact that the principal curvatures of a saddle-like strips can be isometrically switched from positive to negative and vice-versa. This leads to the formation of low energy pathways across the energy barrier corresponding to the flat configuration. 

Furthermore, we find that tristability occurs in strip-like plates when $\det(\boldsymbol{\delta})\sim 0$, where $\boldsymbol{\delta}={\bf D}-{\bf B}{\bf A}^{-1}{\bf B}$ is the effective bending stiffness matrix. In orthotropic materials, this is corresponds to the limit $\beta\rightarrow\nu^{2}$ where $\beta=E_{y}/E_{x}$ discussed by Vidoli \& Maurini (2009) for the case of plates with uniform curvature. A special feature of tristable strips is associated with the height of the barrier separating the two equilibrium configurations of like-sign curvature. The height of this barrier is, in general, much smaller than that associated with the flat configuration. This might have interesting implications for nanoscale materials. Typical strip-like biopolymers, such as proteins or cytoskeletal filaments, have widths of the order of few nanometers, lengths that ranges from 1 to 10$^{3}$ nm and elastic moduli of the order of mega-Pascal. Assuming $\hat{c}_{x}$, $\hat{c}_{y}$ and $\hat{c}_{xy}$ of order one and $at\sim 1$ nm$^2$ one finds that $U(\kappa_{y}^{-})\approx 3\cdot 10^{-4}$ $k_{B}T/{\rm nm}$. Thus even a micron-long strip-like polymer with non-zero spontaneous curvature would be able to fluctuate between minima under the sole effects of thermal excitations. Extending our analysis of multistability to include the effects of thermal fluctuations is likely to yield insights on the dynamical behavior of biopolymer assemblies and is but one extension of our present analysis.
\\[10pt]
{\small {\em Acknowledgments:} we thank the Harvard-NSF MRSEC, the Harvard-Kavli Nano-Bio Science and Technology Center, the Wyss Institute and the MacArthur Foundation for partial support \revision{and the anonymous referees for their detailed suggestions that have improved the paper.}

\end{document}